# Atomistic modeling of a superconductor-transition metal dichalcogenide-superconductor Josephson junction

Jouko Nieminen,[1,2] Sayandip Dhara,[3] Wei-Chi Chiu,[2] Eduardo R. Mucciolo,[3] and Arun Bansil[2]

[1]*Computational Physics Laboratory, Tampere University, FIN-33014 Tampere, Finland*
[2]*Department of Physics, Northeastern University, Boston, Massachusetts, 02115, USA*
[3]*Department of Physics, University of Central Florida, Orlando, Florida, 32816, USA*
(Dated: Version of May 29, 2023)

Using an atomistic tight-binding model, we investigate the characteristics of a Josephson junction formed by monolayers of $MoS_2$ sandwiched between Pb superconducting electrodes. We derive and apply Green's function-based formulation to compute the Josephson current as well as the local density of states in the junction. Our analysis of diagonal and off-diagonal components of the local density of states reveals the presence of triplet superconducting correlations in the $MoS_2$ monolayers and spin-polarized subgap (Andreev bound) states. Our formulation can be extended to other systems where atomistic details and large scales are needed to obtain accurate modeling of Josephson junction physics.

## I. INTRODUCTION

Josephson junctions (JJs), where two superconductor electrodes are separated by a thin barrier layer of an insulating material, continue to provide a fertile ground for the exploration of novel physical phenomena. Much of the recent activity in this area stems from the search for Majorana fermions in connection with topological quantum computing applications [1, 2]. However, other aspects of JJs have also been studied. For instance, it has been recently found that JJs can show rectification effects [3–7]; in addition, when the insulating material is topological, it has been demonstrated experimentally that Josephson currents become $4\pi$-periodic [8–10], corroborating earlier theoretical predictions [11, 12]. JJs have also been used to reveal subtle topological ordering [13] effects. These findings point to the richness of novel possibilities that could be driven by combining the superconductivity of the electrodes with unconventional properties of the barrier layer material.

In this context, few-layer transition-metal dichalcogenides (TMDs) have gained some prominence. Among the TMDs, $MoS_2$ is of particular interest because of its high mobility upon doping or gating. Experiments have shown a high critical current for single- and double-layer $MoS_2$ JJs and strong evidence of multiple Andreev resonances (MARs), which are both suppressed as more layers are added [14, 15]. The homogeneity and tunability of few-layer $MoS_2$ also have the potential to yield high-quality JJs for use in superconducting qubits and superconducting interconnects [16, 17]. Going further, TMDs have been used to fabricate the entire JJ, including the superconductors, yielding high transparency devices [18]. The literature about experimental realizations of TMD-based JJs is already quite robust and it served as motivation for our theoretical modeling effort as atomistic computational studies of these systems are few.

In this work, we develop and employ a formalism for the computation of Josephson currents and local density of states that is suitable for large-scale, atomistic JJ modeling. We apply the formalism to a JJ in which the insulating material consists of an atomically thin layer of $MoS_2$ and the superconductors (SCs) are bulk fcc Pb. $MoS_2$ is chosen as the insulating material due to current interest in fabricating high-quality superconducting qubits with well-controlled functionalized structures, in contrast to traditional $AlO_x$ barriers. As a semiconductor with a relatively wide band gap and strong spin-orbit coupling, the barrier properties of $MoS_2$ are expected to be different from those seen in generic junction models. Pb was chosen as a conventional $s$-type superconductor. We employ a Slater-Koster type multi-band tight-binding model Hamiltonian that accurately reproduces the corresponding first principles band structure. We delineate the role of spin-orbit coupling in $MoS_2$ on the composition of the Josephson current and its dependence on the superconductor phase, as well as on the onset of superconducting correlations in $MoS_2$ via the proximity effect. Our computations show an exponential dependence of the critical current on the number of atomic $MoS_2$ layers, consistent with the experimental data. We find a significant number of subgap MARs for monolayer $MoS_2$, as well as a strong spin polarization and a dependence on the coupling between Pb and $MoS_2$. Spin polarization is weaker in the case of bilayer $MoS_2$, which we attribute to the interplay between the junction's symmetry and spin-orbit coupling (monolayer and bilayer junctions have different symmetry properties). We also identify manifestations of triplet superconducting correlations in the $MoS_2$ layers.

The paper is organized as follows. In Sec. II, we discuss our formulation of the Josephson current (Sec. II A) and the construction of the tight-binding model (Sec. II B). In Sec. III, we present and interpret the results of our numerical calculations. We conclude in Sec. IV with a summary and an outlook. Technical details of the Green's function derivation of Sec. II A are presented in Appendixes A and B. The modeling of the semi-infinite superconducting leads and the details of the tight-binding model are presented in Appendixes C and D, respectively.

## II. METHODOLOGY

This section discusses the methodologies used to obtain the Josephson current and other properties of the SC-TMD-SC junction. We begin with a derivation of the formulation used to compute the current and then describe the construction of the tight-binding model.

### A. Josephson current formulation

There are numerous formulations to compute the dc Josephson current [19–32]. They can be roughly classified into two groups: those based on energy considerations and the explicit contribution of the resonant states at the junction (Andreev bound states), and those based on the integration over a non-equilibrium Green's function that runs across the junction. Here we present an alternative Green's function approach that allows one to separate the superconductor contacts from the normal region, effectively splitting the calculation into two parts. The main advantage of this approach is that only the retarded and advanced Green's functions are needed, which can be efficiently computed using recursive techniques. This method borrows from the quantum-dot literature [33] and, when applied to JJs, allows one to distinguish between the dissipative and coherent contributions. Finite temperature, chemical potential modulations, magnetic fields, microscopic disorder, and spin-orbit couplings can be straightforwardly included in the modeling. In the following, we present the main points in the derivation of an expression for the current that meets the needs for the numerical study presented in this paper; further details are provided in Appendices A and B.

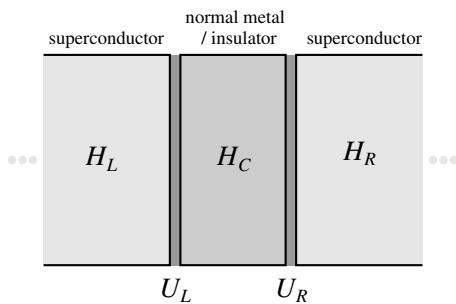

FIG. 1. Schematic representation of a superconductor-normal metal/insulator-superconductor junction. The total Hamiltonian of the system is split into five terms [see Eq. (1)].

We divide the system into two semi-infinite superconductor leads connected to a normal region (see Fig. 1).

The total Hamiltonian of the system is

$$\mathcal{H}_{\text{total}} = \mathcal{H}_L + \mathcal{H}_R + \mathcal{H}_C + \mathcal{U}_L + \mathcal{U}_R, \quad (1)$$

where $\mathcal{H}_{L(R)}$ corresponds to the left (right) superconductors, $\mathcal{H}_C$ corresponds to the finite-size (central) normal region, and $\mathcal{U}_{L(R)}$ denote couplings between the left (right) superconductor and the normal region. To provide explicitly expressions for these terms, we introduce a description of electron operators as four-spinors:

$$\Psi_a = \begin{pmatrix} c_{a\uparrow} \\ c_{a\downarrow} \\ c_{a\downarrow}^\dagger \\ -c_{a\uparrow}^\dagger \end{pmatrix} \quad \text{and} \quad \Psi_a^\dagger = \begin{pmatrix} c_{a\uparrow}^\dagger & c_{a\downarrow}^\dagger & c_{a\downarrow} & -c_{a\uparrow} \end{pmatrix},$$

The anticommuting operators $c_{a\sigma}$ and $c_{a\sigma}^\dagger$ annihilate and create electrons on site $a$ with spin $\sigma$, respectively. The index $a$ denotes a lattice site but can also include additional on-site characteristics such as the atomic orbital number. In terms of the four-spinor operators, the three subsystem contributions to the Hamiltonian can be written (up to a constant) as:

$$\mathcal{H}_l = \frac{1}{2} \sum_{a,a' \in l} \Psi_a^\dagger \hat{H}_{a,a'}^l \Psi_{a'}, \quad (2)$$

where $l = L, R, C$, and

$$\begin{aligned}\hat{H}_{a,a'}^l &= \left[(t_{a,a'} + \delta_{a,a'} v_a)\hat{\sigma}_0 \right.\\&\quad \left.+ i(\lambda_{a,a'}^x \hat{\sigma}_1 + \lambda_{a,a'}^y \hat{\sigma}_2 + \lambda_{a,a'}^z \hat{\sigma}_3)\right]\hat{\tau}_3 \\&\quad + \delta_{a,a'}(\epsilon_x \hat{\sigma}_1 + \epsilon_y \hat{\sigma}_2 + \epsilon_z \hat{\sigma}_3)\hat{\tau}_0 \\&\quad + \delta_{a,a'}\hat{\sigma}_0 \left(\text{Re}\Delta_a \, \hat{\tau}_1 - \text{Im}\Delta_a \, \hat{\tau}_2\right),\end{aligned} \quad (3)$$

where $t_{a,a'}$ are hopping amplitudes, $\lambda_{a,a'}^k$ are spin-orbit coupling constants ($\lambda_{a,a'}^k = -\lambda_{a',a}^k$), and $\epsilon_k$ are Zeeman fields, which may differ for each region of the system, and Re (Im) denote the real (imaginary) part. $\Delta_a$ denotes the local $s$-wave superconductor order parameter in the mean-field approximation. In the normal region, $\Delta_a = 0$. For the couplings between the superconductor and normal regions, we have

$$\mathcal{U}_l = \frac{1}{2} \sum_{a \in l} \sum_{a' \in C} \left( \Psi_a^\dagger \hat{U}_{a,a'}^l \Psi_{a'} + \Psi_{a'}^\dagger \hat{U}_{a',a}^l \Psi_a \right), \quad (4)$$

where $l = R, L$ and

$$\hat{U}_{a,a'}^l = \begin{pmatrix} u_{a,a'}^{l\uparrow\uparrow} & u_{a,a'}^{l\uparrow\downarrow} & 0 & 0 \\ u_{a,a'}^{l\downarrow\uparrow} & u_{a,a'}^{l\downarrow\downarrow} & 0 & 0 \\ 0 & 0 & u_{a,a'}^{l\downarrow\downarrow *} & -u_{a,a'}^{l\downarrow\uparrow *} \\ 0 & 0 & -u_{a,a'}^{l\uparrow\downarrow *} & u_{a,a'}^{l\uparrow\uparrow *} \end{pmatrix}. \quad (5)$$

Here, $u_{a,a'}^{l\sigma\sigma'}$ are the hopping amplitudes between site $a$ in the superconductor $l$ and site $a'$ in the normal region, when the spin orientation goes from $\sigma$ to $\sigma'$ upon hopping. It is convenient to gauge out the superconductor phases from the lead fermionic operators and move them into the coupling operators: Let $\Delta_a = e^{i\phi_l}|\Delta_a|$ for $l = R, L$. After the transformation

$$\Psi'_a = e^{-i\phi_l \hat{\sigma}_0 \hat{\tau}_3/2} \Psi_a \tag{6}$$

we obtain

$$\mathcal{U}_l = \sum_{a \in l} \sum_{a' \in C} \Psi'^{\dagger}_a \hat{\tilde{U}}^l_{a,a'} \Psi_{a'}, \tag{7}$$

where

$$\hat{\tilde{U}}^l_{a,a'} = e^{-i\phi_l \hat{\sigma}_0 \hat{\tau}_3/2} \hat{U}^l_{a,a'}. \tag{8}$$

We can now drop the imaginary part of $\Delta_a$ from the Hamiltonian of Eq. (3) and consider $\Delta_a = |\Delta_a|$. (Hereafter we omit Pauli identity matrices.)

Following the procedure detailed in Appendix A, we arrive at the following expression for the current emanating from the left superconductor:

$$I_L = \frac{2e}{\hbar} \mathrm{Re} \sum_{a \in L} \sum_{a' \in C} \int \frac{d\varepsilon}{2\pi} \mathrm{tr}\left[\hat{\tilde{U}}^L_{a,a'} \hat{\tau}_3 \hat{G}^<_{a',a}(\varepsilon)\right], \tag{9}$$

where the trace runs over spin indices and $\hat{G}^<_{a,a'}(\varepsilon)$ is the lesser (matrix) Green's function in the energy representation. Notice that the Green's function crosses from the left superconductor to the normal region. An analogous expression can be derived for the current emanating from the right superconductor, $L_R$. Since, in the stationary regime, there is no charge accumulation in the central region, $I_L = -I_R$.

In the literature, Eq. (9) is the basis for numerical computations of the Josephson current. Here we use it instead as a starting point for the derivation of an alternative expression that is more practical and suitable for large-scale atomistic modeling. Details of this derivation are provided in Appendix B. The resulting expression is

$$I_L = \frac{e}{2\hbar} \sum_{a,a' \in L} \sum_{a'',a''' \in C} \int \frac{d\varepsilon}{2\pi} \mathrm{tr}\left\{\hat{\tilde{U}}^L_{a,a''}\left[\hat{\tau}_3 \hat{G}^<_{a'',a'''}(\varepsilon) - \hat{G}^<_{a'',a'''}(\varepsilon)\hat{\tau}_3\right]\left(\hat{\tilde{U}}^L_{a''',a'}\right)^{\dagger}\left[\hat{g}^a_{a',a}(\varepsilon) + \hat{g}^r_{a',a}(\varepsilon)\right]\right\}, \tag{10}$$

where $\hat{g}^{r,a}$ are the retarded and advanced surface (matrix) Green's functions defined at the outer, right-most layer of the semi-infinite left superconductor lead *in isolation* (i.e., decoupled from the normal region). The $I_R$ current has an analogous expression.

A remarkable property of Eq. (10) is that each factor within the square brackets is local and can be computed separately in two stages: first, the retarded and advanced surface Green's functions of the decoupled superconductors, which can be computed using decimation techniques [34], as shown in Appendix C; second, the full Green's function of the normal region computed at the left-most surface only. The latter depends implicitly on the coupling to the superconductors and their surface Green's functions, which can be efficiently computed using recursive techniques [35]. It also depends implicitly on the superconductor phases $\phi_L$ and $\phi_R$, although it is only practical to extract an explicit functional dependence in simple situations, such as for one-dimensional chain systems.

We employ Eq. (10) in all numerical calculations of the dc Josephson current presented in this paper.

### B. Tight-binding model

The geometrical structure of the SC-TMD-SC system under study is described in Fig. 2. A supercell was constructed by aligning the zigzag directions of the $MoS_2$ and Pb(111) surfaces. A $2 \times 2$ supercell of $MoS_2$ was set to match a $\sqrt{3} \times \sqrt{3}$ supercell of a Pb(111) electrode. Periodic boundary conditions were applied in the transverse direction ($xy$ plane), while the supercurrent flows in the $z$ direction piercing the two semi-infinite superconducting leads. The superconducting leads are constructed using ABC stacking for an fcc structure, hence the basic building block of the lead has three layers. The method to model semi-infinity is described in Appendix C. In Fig. 2, we show only the single monolayer configuration of $MoS_2$, but we have also studied double-, triple-, and quadruple-layer cases, where the even numbered cases have inversion symmetry while the odd numbered cases have mirror symmetry $M_z$. The stacking of Pb electrodes with fcc(111) orientation follows the symmetry of the $MoS_2$ layer. Hence, for the mirror-symmetric case, the stacking of the top (bottom) electrode follows ABC (CBA) order; for the inversion symmetric case, the top electrode is an image of the bottom one inverted with respect to the inversion point of the $MoS_2$ double layer.

Since the system is periodic in the transverse directions but non-periodic in the longitudinal direction, the calculations are performed using a momentum representation instead of spatial coordinates in the transverse direction. Depending on which quantity is being analyzed, either sums over the horizontal Brillouin zone (or parts of it) are used or quantities in momentum space are Fourier transformed into the simulation cell and projected to an appropriate real-space wave function basis. We note that reciprocal-space lattice vectors of the $2 \times 2$ supercell have half the length of the lattice vectors for the $1 \times 1$ primitive cell of $MoS_2$. This is significant especially in attributing spin-resolved behavior of Andreev bound states to the high-symmetry points of the Brillouin zone (BZ).

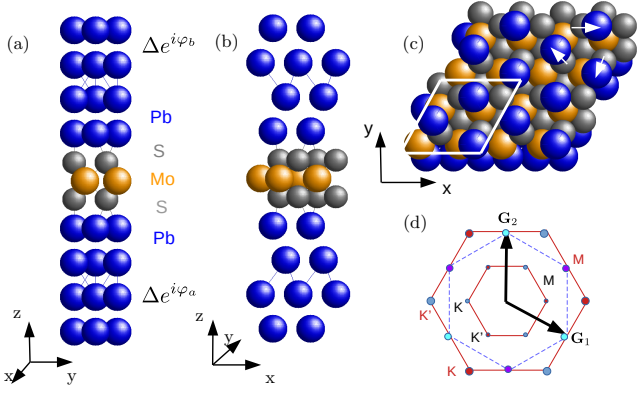

FIG. 2. Geometry of the single-layer Pb-MoS$_2$-Pb junction from two sides: (a) perpendicular to the $yz$ plane and (b) perpendicular to the $xz$ plane. (c) Interface between the top layer of MoS$_2$ and the bottom layer of the Pb electrode. The supercell is indicated by the parallelogram with white boundaries. Notice that there are three Pb atoms and four S atoms in the supercell. Locally, the three Pb atoms are chirally displaced with respect to the three nearest S atoms, while the fourth S atom links to the neighboring supercells. (d) The smallest hexagon is the Brillouin zone (BZ) of the computational $4 \times 4$ supercell, while the largest hexagon shows the BZ of the MoS$_2$ primitive cell. The hexagon with blue dashed lines corresponds to the BZ of the Pb primitive cell, showing the 30° rotation.

| Orbital | $d_{z^2}$ | $d_{xz}$ | $d_{yz}$ | $d_{xy}$ | $d_{x^2-y^2}$ |
|---|---|---|---|---|---|
| $d_{z^2}$ | 0 | $-i\sqrt{3}\sigma_y$ | $i\sqrt{3}\sigma_x$ | 0 | 0 |
| $d_{xz}$ | $i\sqrt{3}\sigma_y$ | 0 | $-i\sigma_z$ | $i\sigma_x$ | $-i\sigma_y$ |
| $d_{yz}$ | $-i\sqrt{3}\sigma_x$ | $i\sigma_z$ | 0 | $-i\sigma_y$ | $-i\sigma_x$ |
| $d_{xy}$ | 0 | $-i\sigma_x$ | $i\sigma_y$ | 0 | $2i\sigma_z$ |
| $d_{x^2-y^2}$ | 0 | $i\sigma_y$ | $i\sigma_x$ | $-2i\sigma_z$ | 0. |

TABLE I. The inter-orbital matrix elements $\mathbf{L}\cdot\boldsymbol{\sigma}$ of $d$ orbitals of Mo for $H_{\mathrm{SOC}}$ according to Ref. [40]. In addition to these terms, there is a common amplitude $\lambda$ chosen to give the correct spin-orbit splitting for bulk MoS$_2$.

The computational basis set consists of $s$, $p_x$, $p_y$, and $p_z$ orbitals for the Pb and S atoms and $s$, $d_{z^2}$, $d_{xz}$, $d_{yz}$, $d_{xy}$, and $d_{x^2-y^2}$ orbitals for the Mo atoms. The hopping integrals follow Slater-Koster form [36, 37] with fitted amplitudes (see AppendixD for details). The electron, hole, and spin degrees of freedom are incorporated into the full Hamiltonian as follows:

$$\mathcal{H} = \sum_{a,b,\sigma} \left( \epsilon_a \, c_{a\sigma}^\dagger c_{a\sigma} + V_{ab} \, c_{a\sigma}^\dagger c_{b\sigma} \right) + \mathcal{H}_{\mathrm{SOC}} + \mathcal{H}_{\mathrm{SC}}. \quad (11)$$

Here, $a, b$ are composite indices that encode both site coordinates and orbital indices within the simulation cell, and the Hamiltonian is Fourier transformed into momentum space in the $x$ and $y$ directions. We use the same parametrization of the tight-binding Hamiltonian as in Refs. [15] and [38] (see also Appendix D). As in Ref. [39], we use a fitted spin-orbit term following Ref. [40],

$$\mathcal{H}_{\mathrm{SOC}} = \sum_{a,\sigma,b,\sigma'} \langle a\sigma | \lambda \mathbf{L} \cdot \boldsymbol{\sigma} | b\sigma' \rangle c_{a\sigma}^\dagger c_{b\sigma'}, \quad (12)$$

which is applied only to the $d$ orbitals of Mo atoms in the present calculations. Table I shows explicitly the on-site matrix elements between the $d$ orbitals of Mo that go into Eq. (12) as derived in Ref. [40]. We choose the value $\lambda = 0.048$ eV for the spin-orbit coupling amplitude.

Following Ref. [15], superconductivity is modeled using the Hamiltonian

$$\mathcal{H}_{\mathrm{SC}} = \sum_{a,b,\sigma} (\Delta_{a,\sigma;b,\bar{\sigma}} c_{a\sigma}^\dagger c_{b\bar{\sigma}}^\dagger + \Delta_{b,\bar{\sigma};a,\sigma}^\dagger c_{b\bar{\sigma}} c_{a\sigma}), \quad (13)$$

where $\Delta_{a,\sigma;b,\bar{\sigma}}$ is the superconductor order parameter. The orbital indices $a$ and $b$ refer to orbitals of Pb atoms; $\bar{\sigma}$ denotes $\sigma$ flipped. For simplicity, $b = a$, and in order to model singlet superconductivity, we choose $\Delta_{a,\uparrow;a,\downarrow} = -\Delta_{a,\downarrow;a,\uparrow}$. The value $\Delta_{a,\uparrow;a,\downarrow} = 1.4$ meV is chosen for calculations according to the experimentally observed gap in Pb [41].

### C. Green's function calculations and visualization of results

The Hamiltonian of Eq. (11) is employed to write Bogoliubov-de Gennes equations and Nambu-Gorkov Green's functions consisting of spin-up and spin-down electron and hole blocks [42],

$$\hat{G}_{\alpha,\beta}(E,\mathbf{k}) = \begin{pmatrix} G_{\alpha,\beta}^{\mathrm{e}}(E,\mathbf{k}) & F_{\alpha,\beta}(E,\mathbf{k}) \\ F_{\alpha,\beta}^\dagger(E,\mathbf{k}) & G_{\alpha,\beta}^{\mathrm{h}}(E,\mathbf{k}) \end{pmatrix},$$

where we use the combined orbital-spin indices $\alpha = a, \sigma$ and $\beta = b, \sigma'$. In the first stage, the Nambu-Gorkov Green's function is calculated in spin and orbital basis as a function of energy and momentum. The resulting function can then be projected in various ways, such as, to spin states, to a path in the momentum space, to a section of the BZ, or partially to the real space.

As noted in Sec. II A, we do not need to determine the Green's function of the entire system to compute properties of the junction. Rather, for the ABS and supercurrent calculations and the analysis of the proximity effect, the noninteracting surface Green's function for the superconducting leads and the full Green's function of the scattering region suffice. In our numerical calculations, the following procedure was employed: first, the noninteracting Green's functions of the leads and the scattering region are calculated separately; second, the recursive method of Appendix C is used to obtain the surface Green's function of the semi-infinite lead; third, self-energies are constructed using

$$\hat{\Sigma}_a^l = \hat{\bar{U}}_{a,a'}^l g_{a'}^l \hat{\bar{U}}_{a',a}^l \quad (14)$$

and used to assemble the full (interacting) Green's function of the scattering region through the solution of Dyson's equation,

$$\hat{G} = \hat{g} + \hat{g}\left(\hat{\Sigma}^L \oplus \hat{\Sigma}^R\right)\hat{G}, \quad (15)$$

where $\hat{g}$ is the non-interacting Green's function and $\hat{G}$ is the resulting full Green's function of the $MoS_2$ part. This interacting Green's function for the scattering region together with the noninteracting surface Green's function of the leads can then be used to calculate the supercurrent in the form of Eq. (10).

In addition to supercurrent calculations, the interacting Green's function of the $MoS_2$ region is used for further analysis of ABSs and the proximity effect. For ABS calculations, the local density of states (LDOS) is obtained from the density matrix

$$\hat{\rho}_{\alpha,\beta}(E,\mathbf{k}) = -\frac{1}{2\pi i}\left[\hat{G}_{\alpha,\beta}(E,\mathbf{k}) - \hat{G}^\dagger_{\alpha,\beta}(E,\mathbf{k})\right]. \quad (16)$$

The density matrix can be projected to chosen sites or orbitals and, specifically, it is used to visualize the LDOS of the superconducting leads and the dependence of the ABSs on the phase difference between the leads. Furthermore, the spin polarization of ABSs and their $\mathbf{k}$ dependence can be analyzed using the density matrix.

The proximity effect and the pairing amplitude within the normal region is studied by taking a real-space projection of the anomalous Green's function component $F_{\alpha,\beta}(E,\mathbf{k})$. Since the spin-orbit coupling tends to mix up and down spins within the normal region and there is a mismatch in spin alignment between the $MoS_2$ layer and the Pb lead, a triplet form of pairing tends to occur in addition to singlet pairing. To illustrate this effect, we arrange the anomalous matrix elements according to the total spin $S = 0, 1$ and $m_S = -S, \ldots, S$ as follows:

$$F^{0,0}_{\alpha\beta}(E,\mathbf{k}) = F_{\uparrow\downarrow}(E,\mathbf{k}) - F_{\downarrow\uparrow}(E,\mathbf{k})$$
$$F^{1,0}_{\alpha\beta}(E,\mathbf{k}) = F_{\uparrow\downarrow}(E,\mathbf{k}) + F_{\downarrow\uparrow}(E,\mathbf{k})$$
$$F^{1,1}_{\alpha\beta}(E,\mathbf{k}) = F_{\uparrow\uparrow}(E,\mathbf{k})$$
$$F^{1,-1}_{\alpha\beta}(E,\mathbf{k}) = F_{\downarrow\downarrow}(E,\mathbf{k}). \quad (17)$$

We then make a Fourier transformation into site space,

$$F^{S,m_S}_{\alpha,\beta}(E,\mathbf{k}) \to F^{S,m_S}_{\alpha,\beta}(E,\mathbf{R}_{\alpha,\beta}). \quad (18)$$

Finally, a real-space projection is performed with Slater-type orbitals,

$$F^{S,m_S}(E,\mathbf{r}) = \sum_{\alpha,\beta} \phi^*_\alpha(\mathbf{r} - \mathbf{R}_\alpha) F^{S,m_S}_{\alpha,\beta}(E,\mathbf{R}_{\alpha,\beta})$$
$$\times \phi_\beta(\mathbf{r} - \mathbf{R}_\beta), \quad (19)$$

where $\mathbf{R}_\alpha$ is the site coordinate of the atom with orbital $\alpha$ and $\mathbf{R}_{\alpha,\beta} = \mathbf{R}_\alpha - \mathbf{R}_\beta$. Furthermore, this can be projected onto cross-sectional planes to allow for the visualization of the real-space distribution of superconductivity pairing within the scattering region and to observe the proportion of induced singlet and triplet superconductivity.

As for the choice of $\mathbf{k}$ points for calculating the supercurrent using Eq. (10), we take the sum over the whole BZ using 1024 points. The same number of $\mathbf{k}$ points are used in calculating the real-space projections of the anomalous Green's function, Eq. (19). The spin-resolved ABSs are calculated using Eq. (16) by summing over 1024 $\mathbf{k}$ points of each quadrant of the BZ.

## III. RESULTS

In the following, we mainly focus on the dependence of the supercurrent $I$ and the ABSs on the phase difference $\varphi$ of the SC order parameter across the $MoS_2$ scattering region under different symmetry conditions and layer thicknesses. Two important factors affect the behavior of $I(\varphi)$ and ABSs at different thicknesses of the $MoS_2$ layers. First, the relatively strong spin-orbit coupling (SOC) may lead to spin-polarized states and facilitate triplet superconductivity. Second, the existence of two non-equivalent valleys at the valence band edge (VBE) of $MoS_2$ combined with time-reversal symmetry leads to a spin flip between the valleys. In the case of an even number of $MoS_2$ layers, no spin-orbit splitting occurs at VBE, whereas for an odd number of layers, the inversion symmetry is broken, which leads to spin-valley coupling [43, 44]. Thus, different behavior of spin polarization of ABSs is expected for monolayer and bilayer $MoS_2$. Moreover, the coupling to Pb leads complicates spin-polarization effects, since the misalignment between the fcc(111)-oriented Pb and the $MoS_2$ surface reduces the symmetry further. Beyond $I(\varphi)$ and ABSs, we consider the proximity effect and the emergence of triplet superconductivity for the $MoS_2$ monolayer and bilayer.

We start by considering ABSs and supercurrents in one-monolayer (one-ML) junctions with $M_z$ mirror symmetry (Fig. 3). Effects of breaking the mirror symmetry by scaling the hopping integrals connecting the orbitals of the $MoS_2$ slab and the upper Pb lead only are also discussed. For this purpose, we follow the parameterization of Ref. [38] and apply the scaling factors 1.00 (full), 0.75 (intermediate), and 0.50 (weak) to the Slater-Koster hopping integrals to the Hamiltonian matrix elements.

The computed supercurrent through the $MoS_2$ monolayer as a function of the phase difference between the Pb leads follows very faithfully a sinusoidal form, which is in agreement with the generic model of a Josephson junction [45–47], where the maximum amplitude occurs at $\varphi = \pi/2$ and $\varphi = 3\pi/2$ [Fig. 3(a)]. As expected, weakening the interaction from the scaling factor of 1.0 to 0.75 and 0.50 lowers the critical current $I_C$ [Fig. 3(b)]. Surprisingly, features of the critical currents for the full and intermediate interactions are seen not to differ significantly. This behavior can be understood by looking at the evolution of the ABSs as a function of phase difference where the in-gap subbands are flatter in the weak-coupling case. In the generic models of superconductor-normal metal-superconductor junctions [47], the supercurrent is related to the dispersion of ABSs as $I(\varphi) \propto \partial E / \partial \varphi$; hence, the flatter the ABS bands, the lower the supercurrent. In this sense, the behavior of $I(\varphi)$ at different coupling values is consistent with the dispersion of in-gap sub-bands, indicating that both the principal ABS bands and the in-gap sub-bands have non-trivial effects on the current.

To set a reference energy window for analyzing the principal and in-gap ABSs within the $MoS_2$ layer, the density of states of the noninteracting superconducting leads is shown in Fig. 3(d). One can see the sharp coherence peaks and the absence of states within the gap as is typical of s-wave superconductivity. In Figs. 3(e)-3(g) we show the phase dependence of the ABSs at different interaction strengths; here, positions of the coherence peaks are marked with black horizontal lines.

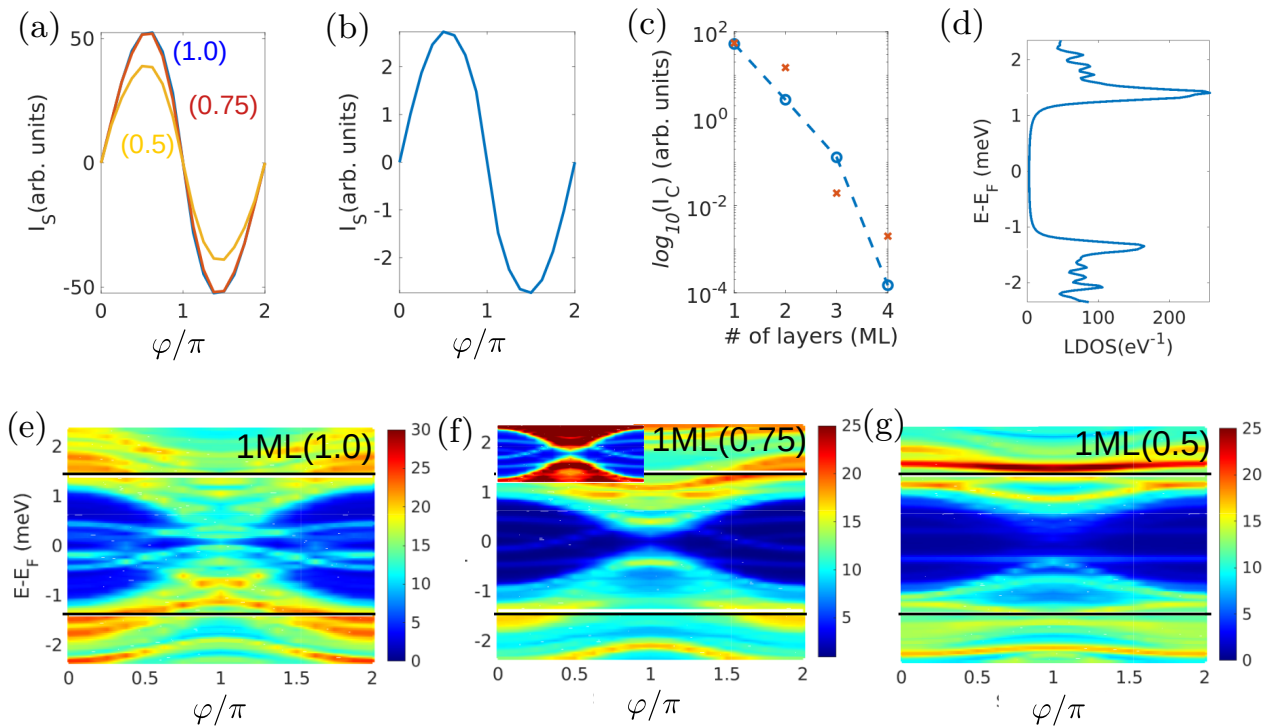

FIG. 3. (a) Supercurrent as a function of the phase difference $\varphi$ between the Pb leads for a one-ML thick $MoS_2$ barrier layer using various interaction strengths between one of the Pb leads and the $MoS_2$ layer. Effects of scaling the Pb-$MoS_2$ overlap integrals are shown. (b) Supercurrent versus phase difference for a bilayer $MoS_2$ (maximum Pb-$MoS_2$ interaction strength used). (c) Dependence of the critical current $I_C$ on the number of $MoS_2$ layers (blue circles) in comparison to experimental results from Ref. [14] (red stars), where MoRe superconducting leads are employed. (d) Density of states of isolated superconducting leads. Evolution of ABSs as a function of the phase difference for (e) full, (f) intermediate, and (g) weak interactions between the upper Pb lead and the $MoS_2$ layer. The inset in (f) highlights the in-gap ABSs.

It is useful to consider the phase dependence of ABSs in terms of their predicted dispersions from a generic weak-link model,

$$E_{\rm gen}(\varphi) = \pm\sqrt{1 - \tau \sin^2(\varphi/2)}, \qquad (20)$$

where $\tau$ is an effective coupling parameter [47–49]. At full Pb-$MoS_2$ interaction strength, the principal ABSs follow the generic behavior closely, nearly crossing at $\varphi = \pi$ [Fig. 3(e)], suggesting an effective coupling parameter $\tau \approx 1$. Weakening of the SC-$MoS_2$ interaction at the upper interface creates a gap between the two Andreev bands at $\varphi = \pi$ [Fig. 3(f) and 3(g)], which is consistent with the prediction for a weak link, i.e., when $0 \leq \tau \ll 1$. Hence, the calculated supercurrent and phase dependence of ABSs both follow very closely the qualitative results of the generic model. However, the detailed structure of ABSs is more complicated than what the expression in Eq. (20) predicts. In the case of a full Pb-$MoS_2$ interaction, a bundle of in-gap states emerges to complicate the overall picture. There appears a split between the upper and lower branches of the nearly flat in-gap bands, which merge when approaching $\varphi = \pi$ from either side. This bundle of in-gap bands is still seen, albeit much weaker and somewhat shifted in energy, in the intermediate interaction case [see inset of Fig. 3(f)]. A possible explanation for the small difference between the $I(\varphi)$ curves between the full and intermediate coupling might be the effective gap separating the main ABS bands due to anti-crossings with the in-gap bands around $\varphi = \pi$.

While the phase dependence of supercurrent and the corresponding ABS bands provide an overall comparison to the generic models, similar calculations for thicker $MoS_2$ layers allow comparison to results on realistic JJs. Increasing the thickness of the $MoS_2$ region leads to an exponential decrease in the critical current, as seen in Fig. 3(c). As a comparison to experiments, we have inserted scaled experimental results for the dependence of the critical current vs. the number of $MoS_2$ layers for a JJ with superconducting MoRe electrodes [14]. We calculated just the critical current for three-ML and four-ML cases, but for a bilayer (two-ML) system, we repeated the same calculations as for the monolayer case. The bilayer system shows an order of magnitude decrease in the critical current but the current versus phase curve follows the same shape as the one-ML case [Fig. 3(b)]. A wide gap in the relatively flat ABS bands of the bilayer case [Fig. 5(a)] consistently correlates a low supercurrent with flat bands resulting from a small value of $\tau$ in the generic models. In addition to low current, the in-gap states are essentially absent apart from a narrow splitting of the principal ABS bands. This point is discussed in connection with the spin polarization of ABSs below.

Next, we discuss the effect of SOC on the $k$-dependent spin polarization of the ABSs. Polarization effects related to the coupling between the electron's spin and momentum may emerge in the case of broken symmetries. A relevant phenomenon for TMDs is spin-valley coupling resulting from SOC and broken inversion symmetry [43, 44, 50]. Momentum-dependent effects may be enhanced if the translational symmetry is also broken, which is the case at edges, surfaces, and interfaces. In a generic form, the SOC interaction term can be expressed as [51],

$$U_{\text{SOC}} \propto \sigma \times \mathbf{p} \cdot \nabla V = \sigma \cdot \mathbf{p} \times \nabla V,$$

where $V$ is the space-dependent potential energy. The parametrized Hamiltonian matrix element of Eq. (12) does not include the gradient of $V$ explicitly, but it captures SOC effects by including directional asymmetry between the interfaces of the junction. Since the system is nonmagnetic and time-reversal symmetry is not broken, summing over the entire BZ hides possible spin dependencies. Therefore, to capture contributions from nonequivalent $K$ and $K'$ valleys, we consider various sections of the BZ as shown in Fig. 4(b).

While a set of in-gap bands emerges in the mirror-symmetric one-ML case where contributions from the entire BZ for the two spin orientations are added together [Fig. 4(a)], a complicated behavior is seen around $\varphi = \pi$ when states are spin-resolved in various BZ quadrants. In panels (c)-(f) of Fig. 4, the spin polarization of the local density of states LDOS, $\rho_\uparrow - \rho_\downarrow$, is plotted for different quadrants of the reciprocal primitive cell [see Fig. 4(b) for the definition of the quadrants]. For the $\mathbf{k}$ points in quadrants Q1 and Q2, the in-gap ABSs are strongly spin-polarized, but they exhibit a more complicated structure than in Fig. 4(a). Careful inspection shows that in addition to the flat in-gap bands, there are two branches that seem extended outside the gap with band crossings near the SC gap edges of the Pb leads. A dramatically different pattern is seen in quadrants Q3 and Q4. Here, the overall dispersion resembles the generic model with a narrow gap at $\varphi = \pi$ with practically no contribution from the in-gap ABSs. Although the upper branches of the principal ABS bands show a notable degree of spin polarization, the overall spin polarization of the in-gap states in quadrants Q3 and Q4 is generally weaker than that in quadrants Q1 and Q2.

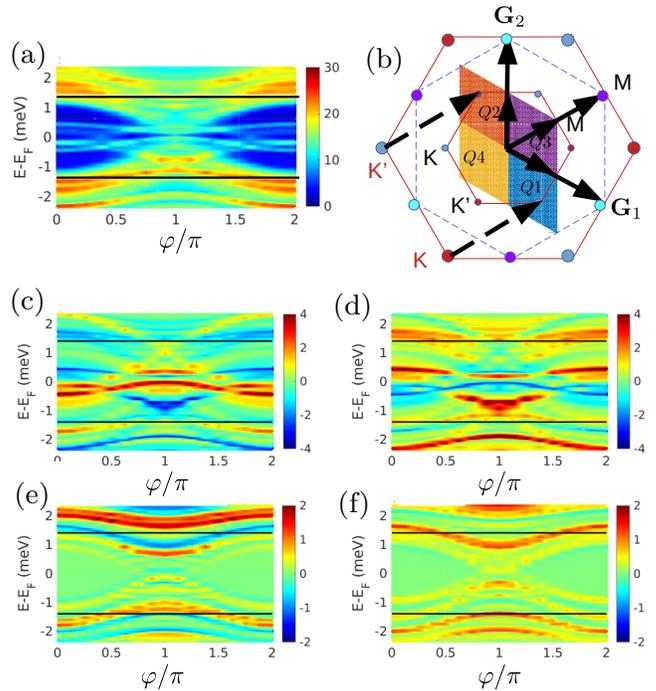

FIG. 4. (a) Local density of states (LDOS) showing ABSs for a single monolayer Pb-$MoS_2$-Pb Josephson junction as a function of the phase difference. (b) The primitive cell of reciprocal space where quadrants are indicated for spin-resolved calculations. Also, the two-dimensional BZ of the system is depicted showing the two non-equivalent valley points $K$ and $K'$, as well as the $M$ point. The inner hexagon is the BZ of the computational supercell ($BZ_s$), and the outer one is the BZ of a $MoS_2$ primitive cell ($BZ_p$). The dashed arrows indicate the folding of $K$ and $K'$ points of $BZ_p$ to the $K'$ and $K$ points of $BZ_s$. (c)-(f) Spin polarization of the LDOS over a quadrant of the primitive cell: (c) quadrant Q1 (blue); (d) quadrant Q2 (red); (e) quadrant Q3 (magenta); and (f) quadrant Q4 (yellow).

Differences in the spin patterns in various quadrants are driven by the spin-valley coupling effects in the TMDs[43, 44, 50]. In honeycomb structures, the $K$ and $K'$ are not equivalent and in the presence of spin-orbit coupling, we obtain spin-polarized states, which are spin flipped in time-reversal symmetric systems. This effect is seen in TMDs with an odd number of layers with broken inversion symmetry, where the band character is strongly $d_{xy}$ and $d_{x^2-y^2}$ at the VBE. Table I shows that the matrix elements between the $d_{xy}$ and $d_{x^2-y^2}$ orbitals have a $\sigma_z$-type coupling that favors spin polarization of the related eigenstates. Furthermore, the geometrical structure of the Pb surface possesses a different symmetry and orientation compared to the horizontal geometry of the $MoS_2$ slabs.

To understand the connection between spin-valley coupling and the spin polarization of the ABSs in various quadrants, note that the $K$ and $K'$ points lie in the middle of quadrants Q1 and Q2, respectively, while the two points corresponding to the $MoS_2$ VBE lie well outside the quadrants Q3 and Q4 [52]. Since a free-hanging $MoS_2$ layer hosts a wide gap between the valence and conduction bands, the Fermi level and thus the superconducting gap lies far from the VBE. Due to the coupling between the orbitals contributing to the VBE and the surface states of the SC leads, however, the proximity-induced states become spin polarized.

Although we find the most significant spin polarization in the direction of $\mathbf{G}_1 - \mathbf{G}_2$ and equivalent directions, this effect may not be observable experimentally in systems that are periodic in the transverse direction since the corresponding $\mathbf{k}$ vectors are perpendicular to the direction of the supercurrent. However, if the lattice is distorted in the horizontal plane, spin-polarized supercurrents may become observable. This would be the case for nanoribbons or components with a finite width.

In contrast to the one-ML case, the two-ML case shows a simpler structure of ABSs. Two factors are responsible for this difference: a thicker barrier for Cooper pairs to tunnel across the scattering region, and the presence of the inversion symmetry instead of the mirror symmetry. Figure 5(a) shows behavior consistent with Eq. (20) with less pronounced features: The inset in Fig. 5(a) shows only a few weak in-gap bands close to the gap edges. The spin-polarized dispersions in the four quadrants of the BZ reveal the spin-valley coupling effects in the vicinity of $K$ or $K'$ outside the gap. The faint branches within the gap regions in panels (b)-(e) of Fig. 5, however, show only small spin polarization. The branches are otherwise symmetric, but the spin polarization is approximately inverted for $\varphi = \pi$. It is plausible that the 30° rotation of the Pb BZ with respect to the $MoS_2$ BZ and the horizontal displacement of the interfacial Pb atoms with respect to S atoms would affect the anomalously small spin polarization here.

Next, we consider the proximity effect within the scattering region by visualizing the anomalous Green's func-tion matrix elements. Panel (a) in Fig. 6 shows a side view of the one-ML junction, and panels (b)-(d) show the real (left) and imaginary (right) parts of the anomalous singlet Green's function $F^{0,0}(E = E_F, \mathbf{r})$ projected into real space and integrated over the $y$ coordinate. For phase difference $\varphi = 0$, the pairing amplitude is predominantly real and follows the horizontal mirror symmetry of the scattering region. A non-zero phase difference between the Pb leads modifies this pattern. For $\varphi = \pi/2$, the real and imaginary parts are equally strong. A phase difference of $\varphi = \pi$ leads to almost real pairing amplitude, but there is a change of sign between the lower and upper layers. As a result, the Mo layer becomes a nodal plane of the singlet pairing amplitude.

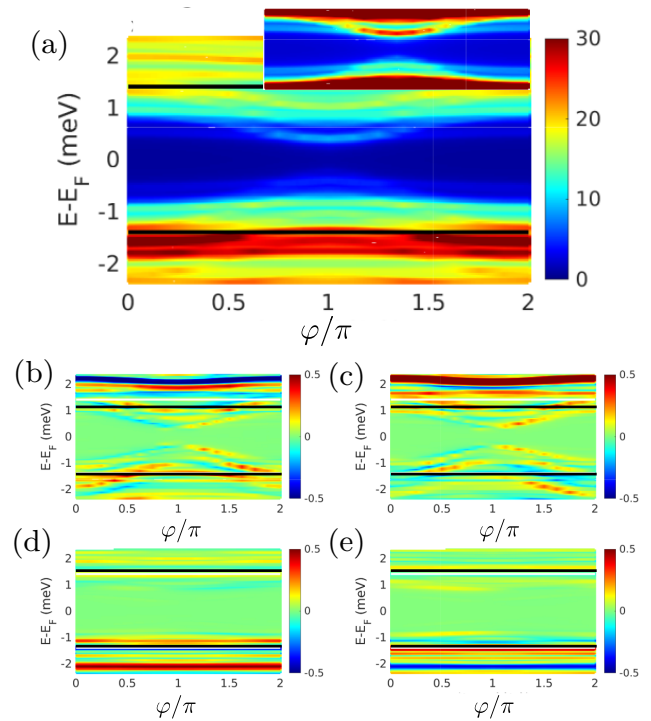

FIG. 5. Similar to Fig. 4 but for a bilayer Pb-$MoS_2$-Pb Josephson junction. The inset in panel (a) highlights the in-gap ABSs close to the gap edges.

We also consider the impact of SOC on the proximity effect and the emergence of triplet pairing. For this purpose, in Figs. 6, 7, and 8 we show the real space projection of the singlet and triplet forms of the anomalous Green's function $F^{S,m_S}$ for the one-ML and two-ML cases. While the Fermi energy in our calculations is in the middle of the $MoS_2$ band gap set by the Pb leads, the effect of spin-valley coupling at VBE is not expected to be strong. Nevertheless, due to the mixing of spin-up and spin-down states and hybridization of the $MoS_2$ and Pb surface states, the emergence of triplet pairing is expected to take place.

Figures 6(b), 6(c), and 7 show that, due to the relatively weak effect of SOC, singlet pairing is stronger

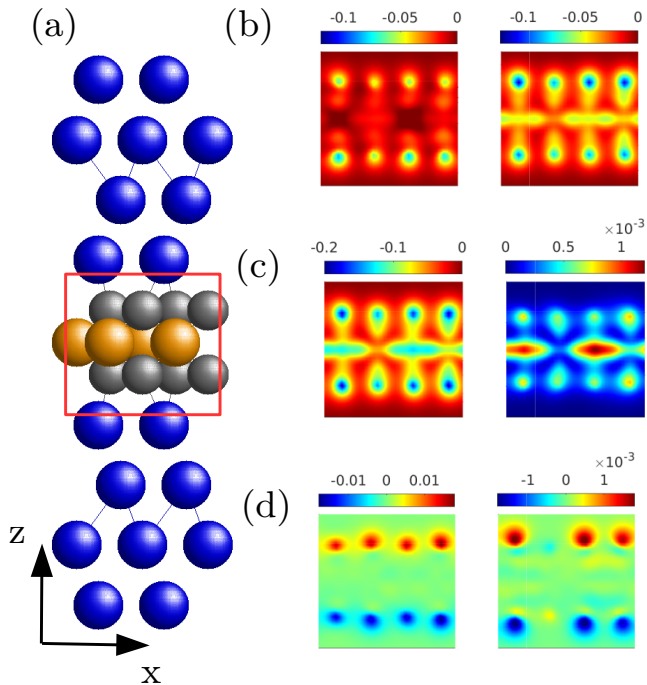

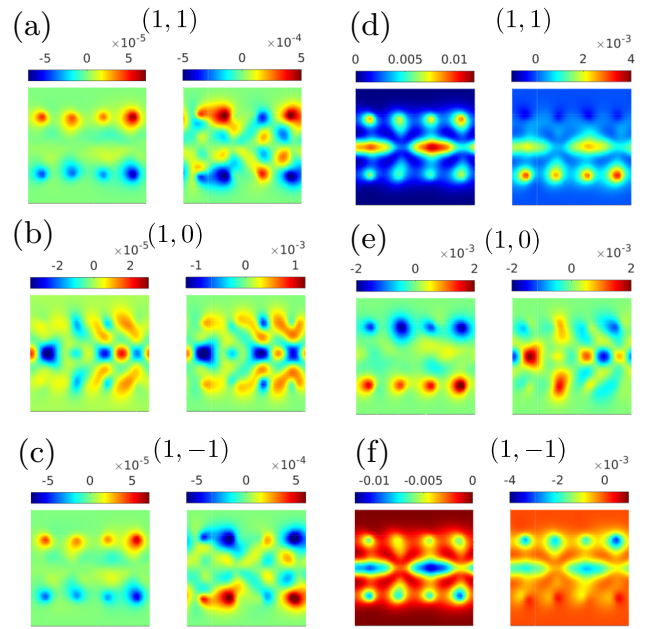

FIG. 6. (a) Atomic arrangement in the $xz$ plane (perpendicular to the junction interfaces) for a monolayer Pb-MoS$_2$-Pb junction. (b)-(d) Real-space projection of the singlet terms of the anomalous matrix elements of the Nambu-Gorkov Green's function $F^{0,0}(E=E_F,\mathbf{r})$ for (b) $\varphi=\pi/2$, (c) $\varphi=0$, and (d) $\varphi=\pi$. The projection corresponds to integration over the $y$ coordinates. The red rectangle in the ball-and-stick graph in (a) indicates the area shown in the projection pictures. Notice the mirror symmetry of the system. The real part of the matrix elements is shown on the left, while the imaginary part is on the right.

FIG. 7. (a)-(f) Real-space projections of the triplet $[(S,m_s) = (1,-1), (1,0), (1,1)]$ terms of the anomalous matrix elements of the Nambu-Gorkov Green's function $F^{S,m_S}(E=E_F,\mathbf{r})$. Plots (a)-(c) correspond to the phase difference $\varphi=0$ between the leads. (d)-(f) correspond to the phase difference $\varphi=\pi/2$. The real part of the matrix elements ise given on the left-hand side figure in each panel, while the imaginary part is given in the right-hand side figures. The atomic arrangement is the same as in Fig. 6(a), where the red rectangle indicates the area shown in the projection pictures.

compared to triplet pairing by roughly two to three orders of magnitude in the one-ML case. For phase difference $\varphi=0$ between the superconducting leads [see Figs. 6(c) and 7(a)-7(c)], both the singlet and the triplet cases with $m_s=0$ follow approximately the mirror symmetry of the system, but there are some notable differences between the patterns. While the singlet projection is essentially real-valued, the triplet case is mainly imaginary. No spatial phase separation between the top and bottom S atomic layers is seen in the $m_s=0$ case, but a clear $p$-wave type change of phase occurs in the triplet $m_s=\pm1$ case with a phase difference of $\pi$ between the top and the bottom of the scattering region. In addition, the spin-flip between $m_s=1$ and $m_s=-1$ is accompanied with a phase inversion of the dominant imaginary part of $F^{S,m_S}$.

The left-hand side figures in panels (d)-(f) of Fig. 7 show the real-space projections of the triplet pairing amplitudes for $\varphi=\pi/2$. For both the singlet [Fig. 6(b)] and the triple case with $m_s=0$, the real and imaginary parts are of the same order of magnitude. The patterns are essentially mirror symmetric, with the exception of the real part of the triplet case, where there is phase inversion between the top and bottom layers. The main difference with respect to the $\varphi=0$ case is in the real and imaginary parts of the patterns, as all the singlet as well as triplet projections are now of the same order of magnitude. For spin-up and spin-down patterns, the dominant real parts are quite mirror-symmetric with phase inversion between the top and bottom layers combined with a spin flip. For the imaginary parts, the mirror symmetry is preserved only if we combine spin-flip and phase inversion.

In Fig. 8, we do not present the projections integrated into the $y$ direction, but the projections in a cross-sectional plane where $y$ is held constant coinciding with a point of inversion symmetry. In the right panel of Fig. 8(a), as an example we show how the S atoms of the upper and lower monolayers transform under the inversion [see Fig. 8(b)], where the real part of the singlet amplitude is seen to be dominant and inversion symmetric. Again, the relatively weak effect of SOC leads to a dominant singlet pairing by roughly two decades. For all three triplet projections, the imaginary part is dominant, and the inversion symmetry holds for the spin-up and spin-down amplitudes. For the triplet $m_s=0$ case, there seems to be slightly more complicated mixing of phases of real and imaginary parts under inversion. A common feature in all cases is that the proximity-induced SC pairing is still relatively strong up to the inner layers of sulfur atoms.

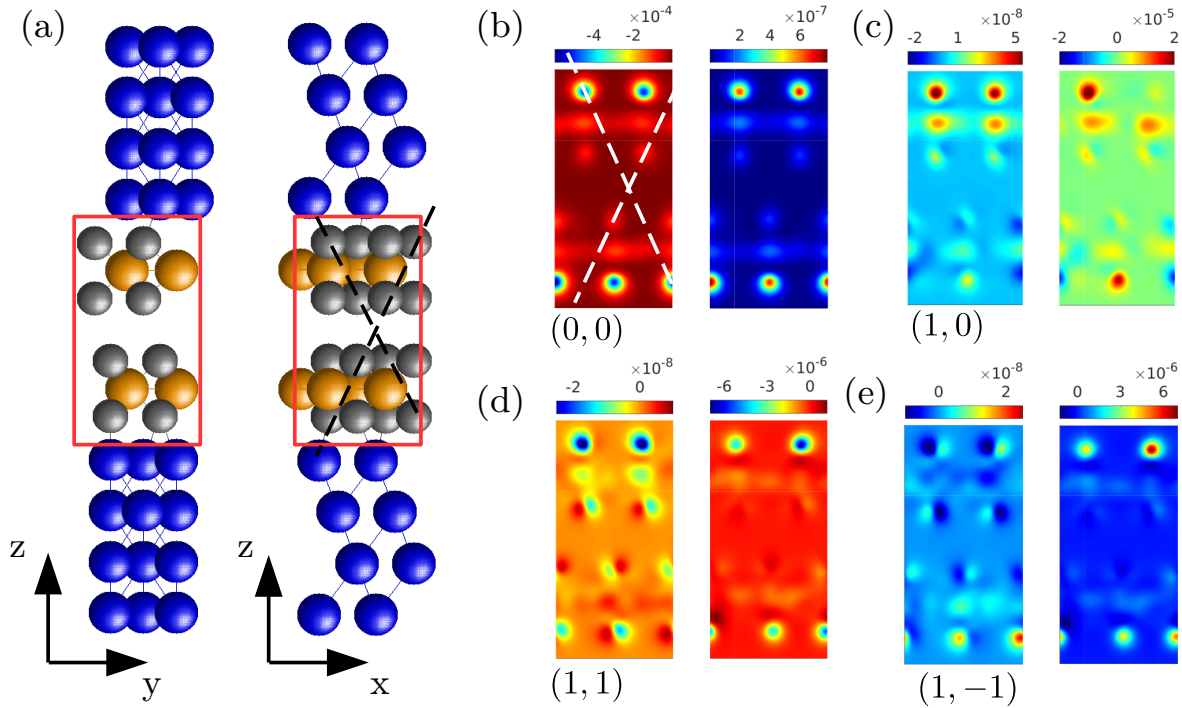

FIG. 8. Similar to Fig. 7, but for a bilayer Pb-MoS$_2$-Pb junction. Only the $\varphi = 0$ case is shown. Notice the inversion symmetry which is indicated by the dashed lines in panels (a) and (b). The amplitudes are projected on a cross-sectional constant-$y$ plane containing the inversion point.

## IV. SUMMARY AND OUTLOOK

We have presented a formalism along with an associated illustrative computational study of Josephson junctions consisting of Pb superconducting leads and layers of MoS$_2$. Using an atomistic tight-binding model to describe Pb and MoS$_2$, we computed superconducting currents, the local density of states, and anomalous components of Green's functions to reveal the composition of resonant (Andreev bound) states in the junction region that contribute to the supercurrent. We have also identified the $k$-dependent spin polarization of these states and their dependence on the Pb-MoS$_2$ coupling strength and the number of atomic MoS$_2$ layers. We also reveal the presence of induced triplet superconductivity in the MoS$_2$ layers.

As expected, the relatively strong spin-orbit coupling in MoS$_2$ induces $k$-dependent spin polarization in the bound states located in the junction region. However, the dispersion and spin texture of these states are rather complex. We break down the contributions to the spin polarization coming from the different areas of the Brillouin zone and show the dominance of states near the $K$ and $K'$ symmetry points, where spin splitting of the valence band of MoS$_2$ is the strongest. We clearly demonstrate the key roles of lattice symmetry and spin-orbit coupling effects in controlling the degree of spin polarization of states: the mirror-symmetric monolayer systems show stronger polarization and a richer structure of in-gap states compared to the inversion-symmetric bilayer systems. This effect also appears in the strength of the induced triplet superconductivity on MoS$_2$, which is markedly stronger in the monolayer systems.

Our study suggests a number of pathways for future studies. One possibility is to engineer spin polarization in Josephson currents by suitably reshaping the contributions from states in different regions of the Brillouin zone. This could be achieved by employing nanoribbons rather than bulk MoS$_2$ layers or using a different junction geometry in order to break the horizontal translation symmetry. Another direction is to employ different materials in the leads to enhance the interplay between lattice symmetry and spin-orbit coupling. Finally, it would be interesting to consider systems where MoS$_2$ is replaced by other topological materials.

The methodology presented in this paper will allow realistic, material-specific modeling and exploration of supercurrents and resonant bound states in more general Josephson-junction-based systems.

## ACKNOWLEDGMENTS

This work was supported by the US Department of Energy (DOE), Office of Science, Basic Energy Sciences Grant No. DE-SC0019275 and it benefited from Northeastern University's Advanced Scientific Computation Center and the National Energy Research Scientific Computing Center through DOE Grant No. DE-AC02-05CH11231, and the resources of the Tampere Center for Scientific Computing (TCSC). The work at Northeastern also benefited from the Massachusetts Technology Collaborative through Award No. 22032.

## Appendix A: Derivation of Eq. (9)

Consider a general electronic Hamiltonian containing both single-particle and mean-field superconductor terms in the form of a tight-binding model,

$$\mathcal{H} = \sum_{a\sigma}\sum_{a'\sigma'} h_{a\sigma,a'\sigma'} c^\dagger_{a\sigma} c_{a'\sigma'} \\ + \sum_a (\Delta_a c^\dagger_{a\uparrow} c^\dagger_{a\downarrow} + \Delta_a^* c_{a\downarrow} c_{a\uparrow}), \quad (A1)$$

where $h_{a\sigma,a'\sigma'} = h^*_{a'\sigma',a\sigma}$ includes both hopping and on-site energy amplitudes and $\Delta_a$ is a local $s$-wave superconductor order parameter. The fermionic operators $c_{a\sigma}$ and $c^\dagger_{a\sigma}$ annihilate and create electrons on site $a$ with spin $\sigma$, respectively. The current emanating from a site $a$ with spin orientation $\sigma$ can be computed from the expectation value of the rate of change of the electron number on that site, namely,

$$I_{a\sigma} = -e\left\langle \frac{d\mathcal{N}_{a\sigma}}{dt} \right\rangle = -\frac{ie}{\hbar}\langle [\mathcal{H}, \mathcal{N}_{a\sigma}] \rangle, \quad (A2)$$

where $\mathcal{N}_{a\sigma} = c^\dagger_{a\sigma} c_{a\sigma}$ and assuming all operators in the Heisenberg picture (for briefness, unless essential, the time dependence is omitted from the operators). The expectation value of the commutator can be readily computed:

$$\langle [\mathcal{H}, \mathcal{N}_{a\sigma}] \rangle = -\sum_{a'\sigma'}(h_{a\sigma,a'\sigma'}\langle c^\dagger_{a\sigma} c_{a'\sigma'} \rangle - h_{a'\sigma',a\sigma}\langle c^\dagger_{a'\sigma'} c_{a\sigma} \rangle) \\ - 2\sum_a (\Delta_a \langle c^\dagger_{a\uparrow} c^\dagger_{a\downarrow} \rangle - \Delta_a^* \langle c_{a\downarrow} c_{a\uparrow} \rangle). \quad (A3)$$

The second contribution on the right-hand side of Eq. (A3) is due to the breaking of particle number conservation. However, when self-consistency is satisfied,

$$\Delta_a^* = -\Lambda \langle c^\dagger_{a\uparrow} c^\dagger_{a\downarrow} \rangle, \quad (A4)$$

and the second contribution vanishes [53–55]. Here $\Lambda$ is an attractive electron-electron interaction coupling constant [56]. In this case, we can write the current as

$$I_{a\sigma} = \frac{ie}{\hbar}\sum_{a'\sigma'}(h_{a\sigma,a'\sigma'}\langle c^\dagger_{a\sigma} c_{a'\sigma'}\rangle - h_{a'\sigma',a\sigma}\langle c^\dagger_{a'\sigma'} c_{a\sigma}\rangle). \quad (A5)$$

An implicit assumption in the derivation of Eq. (A5) is that the breaking of translation invariance near the junction does not modify the self-consistency condition. As shown in Refs. [57] and [58], such a modification may occur, leading to corrections to the critical current and the current-phase relationship. However, Ref. [57] also argues that these corrections are relatively small for the current-phase relation, which is the main focus of our study. Accordingly, we have neglected these corrections here.

Equation (A5) can be recast in terms of non-equilibrium Green's functions: consider the lesser Green's function

$$G^<_{a\sigma,a'\sigma'}(t,t') \equiv i\langle c^\dagger_{a'\sigma'}(t') c_{a\sigma}(t)\rangle, \quad (A6)$$

where we make explicit the time dependencies of the operators. Using Eq. (A6), we can rewrite the current as

$$I_{a\sigma}(t) = \frac{e}{\hbar} \sum_{a'\sigma'} [h_{a\sigma,a'\sigma'} G^<_{a'\sigma',a\sigma}(t,t) - h_{a'\sigma',a\sigma} G^<_{a\sigma,a'\sigma'}(t,t)]. \quad (A7)$$

Since $G^<_{m,n}(t,t') = -[G^<_{n,m}(t',t)]^*$, we obtain

$$I_{a\sigma}(t) = \frac{2e}{\hbar} \mathrm{Re} \sum_{a'\sigma'} [h_{a\sigma,a'\sigma'} G^<_{a'\sigma',a\sigma}(t,t)], \quad (A8)$$

where Re stands for the real part. Equation (A8) is the basis for the next step.

Once we split the system into three regions as shown in Fig. 1, we can use Eq. (A8) to express the charge current flowing from the left-hand side superconductor into the normal region as

$$I_L(t) = \frac{2e}{\hbar} \mathrm{Re} \sum_{a \in L} \sum_{a' \in C} \sum_{\sigma,\sigma'} \mathrm{tr} \left[ \hat{\bar{U}}^l_{a,a'} \hat{\tau}_3 \hat{G}^<_{a',a}(t,t) \right], \quad (A9)$$

where we have reverted to the four-spinor representation of the coupling matrices and Green's function, namely,

$$\left[ \hat{G}^<_{a',a}(t',t) \right]_{j',j} \equiv i \langle [\Psi^\dagger_{a'}(t')]_{j'} [\Psi_a(t)]_j \rangle. \quad (A10)$$

It is straightforward to derive an expression similar to Eq. (A9) for the current emanating from the right-hand side superconductor. Since we are focused on the zero-bias dc current, which is stationary and thus time independent, we can switch to an energy or frequency representation, in which case the expression for the left current can be written as

$$I_L = \frac{2e}{\hbar} \mathrm{Re} \sum_{a \in L} \sum_{a' \in C} \sum_{\sigma,\sigma'} \int \frac{d\varepsilon}{2\pi} \mathrm{tr} \left[ \hat{\bar{U}}^L_{a,a'} \hat{\tau}_3 \hat{G}^<_{a',a}(\varepsilon) \right], \quad (A11)$$

where

$$\hat{G}^<(t,t') = \hat{G}^<(t-t') = \int \frac{d\varepsilon}{2\pi} e^{-i\varepsilon(t-t')/\hbar} \hat{\tilde{G}}^<(\varepsilon). \quad (A12)$$

(Hereafter we drop the tilde from the Green's function in the energy representation.) It is possible to write the current entirely in terms of equilibrium Green's functions by employing the fluctuation-dissipation theorem [59], which in this context reads

$$\hat{G}^<(\varepsilon) = f(\varepsilon) \left[ \hat{G}^a(\varepsilon) - \hat{G}^r(\varepsilon) \right], \quad (A13)$$

where $f(\varepsilon)$ is the Fermi-Dirac distribution and $\hat{G}^{r(a)}(\varepsilon)$ is the retarded (advanced) Green's function.

### Appendix B: Derivation of Eq. (10)

Let us introduce the time-ordered Green's function

$$\left[ \hat{G}^t_{a,a'}(t-t') \right]_{j,j'} = -i \left\langle T \left\{ [\Psi_a(t)]_j [\Psi^\dagger_{a'}(t')]_{j'} \right\} \right\rangle, \quad (B1)$$

where $T$ denotes time ordering over the Keldysh contour. We assume $a \in C$ and $a' \in L$. To make the notation more compact, we lump indices into a single one, namely, $a, j \to \alpha$ and $a', j' \to \alpha'$. Notice that the domain of $\alpha$ ($L$, $R$, or $C$) can often be inferred from the Green's functions and the interaction factors surrounding them (e.g., $U_L$, $U_R$, $H_C$, etc.).

The equation of motion for the time-ordered Green's function is

$$-i\hbar \frac{\partial}{\partial t'} G^t_{\alpha,\alpha'}(t-t') = -\frac{1}{\hbar} \left\langle T \left\{ \Psi_\alpha(t) \frac{d}{dt'} \Psi^\dagger_{\alpha'}(t') \right\} \right\rangle$$
$$= -i \langle T \{ \Psi_\alpha(t) [\mathcal{H}, \Psi^\dagger_{\alpha'}(t')] \} \rangle$$
$$= -i \langle T \{ \Psi_\alpha(t) [\mathcal{H}_L, \Psi^\dagger_{\alpha'}(t')] \} \rangle$$
$$- i \langle T \{ \Psi_\alpha(t) [\mathcal{U}_L, \psi^\dagger_{\alpha'}(t')] \} \rangle. \quad (B2)$$

We can readily compute the two commutators ($\alpha' \in L$):

$$[\mathcal{H}_L, \Psi^\dagger_{\alpha'}(t')] = \sum_{\alpha \in L} \Psi^\dagger_\alpha(t') H^L_{\alpha,\alpha'} \quad (B3)$$

and

$$[\mathcal{U}_L, \Psi^\dagger_{\alpha'}(t')] = \sum_{\alpha'' \in C} \Psi^\dagger_{\alpha''}(t') \bar{U}^L_{\alpha'',\alpha'}.$$

Inserting the above results into Eq. (B2), we get

$$-i\hbar \frac{\partial}{\partial t'} G^t_{\alpha,\alpha'}(t-t') = \sum_{\alpha'' \in L} G^t_{\alpha,\alpha''}(t-t') H^L_{\alpha'',\alpha'}$$
$$+ \sum_{\alpha''' \in C} G^t_{\alpha,\alpha'''}(t-t') \bar{U}^L_{\alpha''',\alpha'},$$

which can be written more concisely as

$$\left[ G^t_{\alpha,\alpha'} \left( g^t \right)^{-1} \right] (t-t') = \sum_{\alpha''' \in C} G^t_{\alpha,\alpha'''}(t-t') \bar{U}^L_{\alpha''',\alpha'}, \quad (B4)$$

where $g^t$ is the time-ordered Green's function of the left superconductor in isolation, namely, when $\bar{U}^L = 0$. Operating on this equation with $g^t$ from the right, we obtain

$$G^t_{\alpha,\alpha'}(t-t') = \sum_{\alpha'' \in L} \sum_{\alpha''' \in C} \int dt_1 G^t_{\alpha,\alpha'''}(t-t_1) \bar{U}^L_{\alpha''',\alpha'}$$
$$\times g^t_{\alpha'',\alpha'}(t_1 - t'), \quad (B5)$$

or, more compactly,

$$G^t(t-t') = \int dt_1 G^t(t-t_1) \bar{U}^L g^t(t_1 - t'), \quad (B6)$$

where all matrix element summations have turned into matrix multiplications. This equation is now in a form suitable for the Langreth rule [33]

$$C = \int_{\substack{\text{complex}\\\text{contour}}} A\,B \to C^< = \int_{\substack{\text{real}\\\text{time}}} \left[ A^r B^< + A^< B^a \right],$$

which amounts to an analytical continuation into the real axis, leading to

$$G^<(t-t') = \int dt_1 \left[ G^r(t-t_1) \bar{U}^L g^<(t_1-t') + G^<(t-t_1) \bar{U}^L g^a(t_1-t') \right]. \quad (B7)$$

Switching to a frequency representation, we find

$$G^<(\varepsilon) = \left[ G^r(\varepsilon) \bar{U}^L g^<(\varepsilon) + G^<(\omega) \bar{U}^L g^a(\varepsilon) \right]. \quad (B8)$$

Finally, substituting this result into Eq. (9), we obtain the following expression for the stationary current:

$$I_L = \frac{2e}{\hbar} \text{Re} \int \frac{d\varepsilon}{2\pi} \text{Tr} \left\{ \bar{U}^L \hat{\tau}_3 \left[ G^r(\varepsilon) \bar{U}^L g^<(\varepsilon) + G^<(\varepsilon) \bar{U}^L g^a(\varepsilon) \right] \right\}. \quad (B9)$$

The trace acts on the augmented site space. Notice that the Green's functions with capital letters correspond to the normal region, while the ones in lowercase are for the left superconductor in isolation.

At this point, it is useful to represent the lead Green's function $g$ in its energy eigenbasis,

$$[g(\varepsilon)]_{\alpha'',\alpha} = \sum_\kappa (O^{-1})_{\alpha'',\kappa} [g_d(\varepsilon)]_\kappa O_{\kappa,\alpha}, \quad (B10)$$

which leads to

$$I_L = \frac{2e}{\hbar} \text{Re} \int \frac{d\varepsilon}{2\pi} \text{Tr} \left\{ \hat{\tau}_3 \left[ G^r(\varepsilon) \bar{U}_h^L g_d^<(\varepsilon) \bar{U}_h^L + G^<(\varepsilon) \bar{U}_h^L g_d^a(\varepsilon) \bar{U}_h^L \right] \right\}, \quad (B11)$$

where we introduced the hybrid coupling matrices

$$[\bar{U}_h^L]_{\kappa;\alpha'} = \sum_{\alpha \in L} O_{\kappa,\alpha} [\bar{U}^L]_{\alpha,\alpha'}$$

$$[\bar{U}_h^L]_{\alpha''';\kappa} = \sum_{\alpha'' \in L} [\bar{U}^L]_{\alpha''',\alpha''} (O^{-1})_{\alpha'',\kappa}$$

and added a subscript "d" to differentiate the lead's Green's function in the eigenenergy basis representation (where it is diagonal) from the one in the site basis. In fact, when expressed in the energy eigenbasis, the lead's Green's function depends only on the energy eigenvalues $\{\varepsilon_\kappa\}$, namely,

$$[g_d^<(\varepsilon)]_\kappa = 2\pi i f_L(\varepsilon) \delta(\varepsilon - \varepsilon_\kappa) \quad (B12)$$

and

$$[g_d^a(\varepsilon)]_\kappa = [\varepsilon - \varepsilon_\kappa - i0^+]^{-1}. \quad (B13)$$

Using these expressions, we can make further progress. Consider the first term within the curly brackets on the right-hand side of Eq. (B11):

$$\text{Tr} \left\{ \hat{\tau}_3 \left[ G^r(\varepsilon) \bar{U}_h^L g_d^<(\varepsilon) \bar{U}_h^L \right] \right\} = i f_L(\varepsilon) \text{Tr} \left[ \hat{\tau}_3 G^r(\varepsilon) \Gamma_L(\varepsilon) \right], \quad (B14)$$

where we introduced the level width matrix

$$[\Gamma_L(\varepsilon)]_{\alpha''',\alpha'} = 2\pi \sum_\kappa \delta(\varepsilon - \varepsilon_\kappa) [\bar{U}_h^L]_{\alpha''';\kappa} [\bar{U}_h^L]_{\kappa;\alpha'}. \quad (B15)$$

Moreover, notice that

$$\mathcal{R} \left\{ i f_L(\varepsilon) \text{Tr} \left[ \hat{\tau}_3 G^r(\varepsilon) \Gamma_L(\varepsilon) \right] \right\} = \frac{i}{2} f_L(\varepsilon) \text{Tr} \left\{ \Gamma_L(\varepsilon) \left[ \hat{\tau}_3 G^r(\varepsilon) - G^a(\varepsilon) \hat{\tau}_3 \right] \right\}, \quad (B16)$$

We can also manipulate the second term on the right-hand side of Eq. (B11) to obtain a similarly compact form:

$$\text{Re} \left\{ \text{Tr} \left[ \hat{\tau}_3 G^< \bar{U}_h^L g_d^a \bar{U}_h^L \right] \right\} = \frac{1}{4} \left\{ i \text{Tr} \left[ (\hat{\tau}_3 G^< + G^< \hat{\tau}_3) \Gamma_L \right] + \text{Tr} \left[ (\hat{\tau}_3 G^< - G^< \hat{\tau}_3) \bar{U}_h^L (g_d^a + g_d^r) \bar{U}_h^L \right] \right\}. \quad (B17)$$

If we combine the first term inside the curly brackets on the right-hand side of Eq. (B17) with the expression in Eq. (B16), we obtain the following contribution to the current:

$$I_L^{(1)} = \frac{ie}{\hbar} \int \frac{d\varepsilon}{2\pi} \text{Tr} \left( \Gamma_L(\varepsilon) \left\{ \frac{1}{2} \left[ \hat{\tau}_3 G^<(\varepsilon) + G^<(\varepsilon) \hat{\tau}_3 \right] \right\} + f_L(\varepsilon) \left[ \hat{\tau}_3 G^r(\varepsilon) - G^a(\varepsilon) \hat{\tau}_3 \right] \right). \quad (B18)$$

An equivalent exercise for the right lead's current yields

$$I_R^{(1)} = \frac{ie}{\hbar} \int \frac{d\varepsilon}{2\pi} \text{Tr} \left( \Gamma_R(\varepsilon) \left\{ \frac{1}{2} \left[ \hat{\tau}_3 G^<(\varepsilon) + G^<(\varepsilon) \hat{\tau}_3 \right] \right\} + f_R(\varepsilon) \left[ \hat{\tau}_3 G^r(\varepsilon) - G^a(\varepsilon) \hat{\tau}_3 \right] \right). \quad (B19)$$

Since there is no charge accumulation in the system in the stationary regime, we expect $I_L = -I_R \equiv I$. We can symmetrize the left-to-right current by writing $I = (I_L - I_R)/2$ and then combine it with Eqs. (B18) and (B19) to obtain

$$I^{(1)} = \frac{ie}{2\hbar} \int \frac{d\varepsilon}{2\pi} \text{Tr} \left\{ \frac{1}{2}[\Gamma_L(\varepsilon) - \Gamma_R(\varepsilon)] \left[\hat{\tau}_3 G^<(\varepsilon) + G^<(\varepsilon)\hat{\tau}_3\right] + [f_L(\varepsilon)\Gamma_L(\varepsilon) - f_R(\varepsilon)\Gamma_R(\varepsilon)] \left[\hat{\tau}_3 G^{\text{r}}(\varepsilon) - G^{\text{a}}(\varepsilon)\hat{\tau}_3\right] \right\}. \tag{B20}$$

At zero bias, assuming the same temperature in both superconductors, we can write $f_R(\varepsilon) = f_L(\varepsilon) = f(\varepsilon)$. Moreover, in equilibrium, $G^<(\varepsilon) = f(\varepsilon)[G^{\text{a}} - G^{\text{r}}]$, yielding

$$I^{(1)} = \frac{ie}{4\hbar} \int \frac{d\varepsilon}{2\pi} f(\varepsilon) \text{Tr} \left\{ [\Gamma_L(\varepsilon) - \Gamma_R(\varepsilon)] \left[\hat{\tau}_3, G^{\text{a}}(\varepsilon) + G^{\text{rxs}}(\varepsilon)\right] \right\}. \tag{B21}$$

Clearly, when the superconductor leads are identical, this contribution to the current vanishes (it can be shown that the level width matrices $\Gamma_{R,L}$ do not depend on the superconductor phases). When the leads are not identical, consider

$$G^{\text{a}}(\varepsilon) + G^{\text{r}}(\varepsilon) = \hat{\tau}_0 \, G_1(\varepsilon) + \hat{\tau}_1 \, G_1(\varepsilon) + \hat{\tau}_2 \, G_2(\varepsilon) + \hat{\tau}_3 \, G_3(\varepsilon). \tag{B22}$$

Then,

$$[\hat{\tau}_3, G^{\text{a}}(\varepsilon) + G^{\text{r}}(\varepsilon)] = -2i\,\hat{\tau}_2\,G_1(\varepsilon) + 2i\,\hat{\tau}_1\,G_2(\varepsilon) \tag{B23}$$

which can only be nonzero if the $\Gamma_{R,L}$ matrices allow for particle-hole conversion. In the absence of such a conversion, the trace vanishes in Eq. (B21) and $I^{(1)} = 0$.

Let us now return to the contribution to the current originating from the second term on the right-hand side of Eq. (B17). Switching back to the site-only representation for the coupling matrices and to surface Green's functions,

$$\begin{aligned} I_L^{(2)} &= \frac{e}{2\hbar} \int \frac{d\varepsilon}{2\pi} \text{Tr} \left\{ [\hat{\tau}_3, G^<(\varepsilon)] \bar{U}_h^L \left[g_{\text{d}}^{\text{a}}(\varepsilon) + g_{\text{d}}^{\text{r}}(\varepsilon)\right] \bar{U}_h^L \right\} \\ &= \frac{e}{2\hbar} \int \frac{d\varepsilon}{2\pi} \text{Tr} \left\{ [\hat{\tau}_3, G^<(\varepsilon)] \bar{U}^L \left[g^{\text{a}}(\varepsilon) + g^{\text{r}}(\varepsilon)\right] \bar{U}^L \right\}, \end{aligned} \tag{B24}$$

we obtain Eq. (10). The current coming from the right superconductor has an analogous expression.

**Appendix C: Modeling of the semi-infinite leads**

The Green's function of the semi-infinite leads is modeled by applying Dyson's equation recursively. The method is illustrated in Fig. 9. We start with a noninteracting slab with a minimal set of layers in the $z$ direction with $n_s$ orbitals. A Green's function $n_s \times n_s$ matrix $G^{(1)}$ of an isolated slab is calculated and a coupling matrix $t$ between the two consecutive slabs is formed. From these two, a two-slice decoupled Green's function

$$g^{(2)} = G^{(1)} \oplus G^{(1)} = \begin{pmatrix} G^{(1)} & 0 \\ 0 & G^{(1)} \end{pmatrix} \tag{C1}$$

and the corresponding interaction matrix

$$v^{(2)} = \begin{pmatrix} 0 & t \\ t & 0 \end{pmatrix} \tag{C2}$$

are formed. After that, Dyson's equation

$$G^{(2)} = g^{(2)} + g^{(2)} v^{(2)} G^{(2)} \tag{C3}$$

is applied to form the coupled two-slab Green's function. This stage gives us a starting point for a quickly converging iteration, where we construct four-slab Green's functions from the two-slab Green's functions.

In the first step of iteration, we create an initial four-slab Green's function

$$G_0^{(4)} = G^{(2)} \oplus G^{(2)} = \begin{pmatrix} G^{(2)} & 0_{2\times 2} \\ 0_{2\times 2} & G^{(2)} \end{pmatrix} \tag{C4}$$

and the coupling matrix

$$v^{(4)} = \begin{pmatrix} 0 & 0 & 0 & 0 \\ 0 & 0 & t & 0 \\ 0 & t & 0 & 0 \\ 0 & 0 & 0 & 0 \end{pmatrix}. \tag{C5}$$

In the iteration steps we need only four blocks of the previous four-slab Green's functions, as follows:

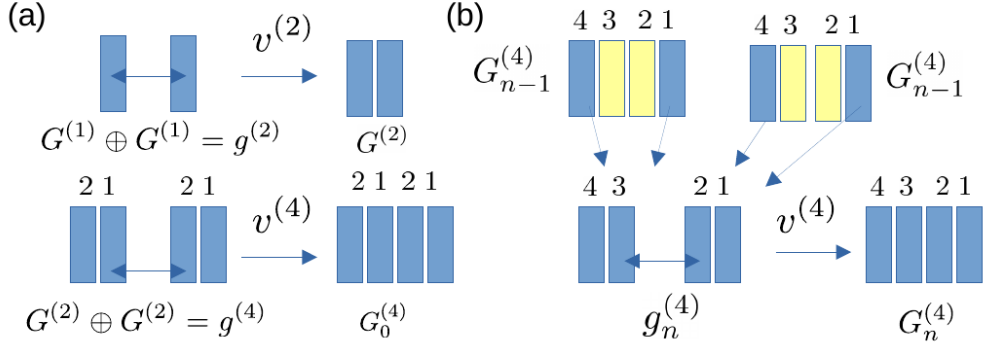

FIG. 9. Description of the recursion method. (a) First, a two-slab decoupled Green's function is created from the one-slab Green's function, then Dyson's equation is applied to obtain the coupled two-slab Green's function. The resulting two-slab Green's functions are used to create the decoupled four-slab Green's function. Dyson's equation is used to get the initial four-slab Green's function. (b) In the insertion process, the two middle slabs are removed to form new two-slab Green's functions, which are combined to a new decoupled four-slab Green's function. After that, Dyson's equation is applied to get the next four-slab Green's function. In calculating Andreev bound states and supercurrents, only the matrix elements of the slab closest to the normal region are necessary.

$$g_n^{(4)} = \begin{pmatrix} G_{n-1}^{(1)}(1,1) & G_{n-1}^{(1)}(1,4) & 0 & 0 \\ G_{n-1}^{(1)}(4,1) & G_{n-1}^{(1)}(1,1) & 0 & 0 \\ 0 & 0 & G_{n-1}^{(1)}(1,1) & G_{n-1}^{(1)}(1,4) \\ 0 & 0 & G_{n-1}^{(1)}(4,1) & G_{n-1}^{(1)}(1,1) \end{pmatrix}, \tag{C6}$$

where $G_{n-1}^{(1)}(i,j)$ is a block consisting of Green's function matrix elements of the $(n-1)$th iteration between slabs $i$ and $j$. The next iterated Green's function is obtained using Dyson's equation

$$G_n^{(4)} = g_n^{(4)} + g_n^{(4)} v^{(4)} G_n^{(4)}. \tag{C7}$$

Finally, we use the block $G_n^{(4)}(1,1)$ as the surface Green's function.

The method converges very fast. If there are $n_s$ orbitals in a slab, the size of the lead after $n$ iterations is $2 \times 2^n = 2^{n+1}$.

### Appendix D: Tight-binding model parameters

The overlap integrals between the atomic orbitals of two atoms can be split into two parts: an angular part that depends on the orientation of the two orbitals, and an amplitude part that depends on the type of bonding between the orbitals ($\sigma$, $\pi$, or $\delta$). We use the same parametrization as in Refs. [39], [15] and [38] (see tables II and III). In Ref. [39], the parameter fitting for MoS$_2$ was done to follow the *abinitio* band structure with special emphasis on the top of the valence bands (TVB) and on the bottom of the conduction bands. In addition, the evolution of the TVB at the $K$ point for film thicknesses one ML to three ML is captured correctly. In Ref. [15], the connection to a Pb substrate was constructed based on the *abinitio* band structure of Pb and the band structure of 1one-ML MoS2 on a Pb(111) slab. The ab-initio band structure of Pb near the Fermi energy is rather complicated, and the small orbital basis used can only follow some of the most salient band features. The interaction between the $s$ and $p$ orbitals of Pb and the adjacent S layer is based on Slater-Koster hopping integrals. The amplitudes are fitted so that the metallicity-inducing in-gap states are reproduced in reasonable agreement with the corresponding *abinitio* calculations.

| Amplitudes | $\varepsilon_s$(Pb) | $\varepsilon_p$(Pb) | $\varepsilon_s$(S) | $\varepsilon_p$(S) | $\varepsilon_s$(Mo) | $\varepsilon_d$(Mo)) |
|---|---|---|---|---|---|---|
| $ss\sigma$ | -6.150 | 0.800 | -10.472 | -3.172 | -21.472 | -1.272 |
| $sp\sigma$ | 3.6878 | 3.6878 | 3.2481 | 3.2481 | 3.2481 | 3.2481 |

TABLE II. On-site energies and the cutoff ranges for the tight-binding Hamiltonian.

| Amplitudes | Pb-Pb | Pb-S | S-S | S-Mo | Mo-Mo |
|---|---|---|---|---|---|
| $ss\sigma$ | -0.90 | -1.20 | -1.60 | -1.82 | -2.08 |
| $sp\sigma$ | 0.94 | 0.71 | 0.53 | 1.10 | |
| $pp\sigma$ | 2.19 | 1.43 | 0.53 | | |
| $pp\pi$ | -0.81 | -0.10 | -0.01 | | |
| $sd\sigma$ | | | | -3.94 | -3.01 |
| $pd\sigma$ | | | | -4.30 | |
| $pd\pi$ | | | | -2.40 | |
| $dd\sigma$ | | | | | -20.8 |
| $dd\pi$ | | | | | 12.0 |
| $dd\delta$ | | | | | -1.77 |

TABLE III. Amplitudes of the Slater-Koster hopping parameters used in the tight-binding Hamiltonian.